# Spin torque gate magnetic field sensor

Hang Xie, Xin Chen, Ziyan Luo, and Yihong Wu[*]

Department of Electrical and Computer Engineering, National University of Singapore,

Singapore 117583, Singapore

Spin-orbit torque provides an efficient pathway to manipulate the magnetic state and magnetization dynamics of magnetic materials, which is crucial for energy-efficient operation of a variety of spintronic devices such as magnetic memory, logic, oscillator, and neuromorphic computing. Here, we describe and experimentally demonstrate a strategy for the realization of a spin torque gate magnetic field sensor with extremely simple structure by exploiting the longitudinal field dependence of the spin torque driven magnetization switching. Unlike most magnetoresistance sensors which require a delicate magnetic bias to achieve a linear response to the external field, the spin torque gate sensor can achieve the same without any magnetic bias, which greatly simplifies the sensor structure. Furthermore, by driving the sensor using an ac current, the dc offset is automatically suppressed, which eliminates the need for a bridge or compensation circuit. We verify the concept using the newly developed $WTe_2$/Ti/CoFeB trilayer and demonstrate that the sensor can work linearly in the range of ±3-10 Oe with negligible dc offset.

[*] E-mail: elewuyh@nus.edu.sg

When a charge current passes through a ferromagnet (FM) / heavy metal (HM) bilayer, spin-orbit torque (SOT) is induced which exerts on the magnetization of the FM layer[1-6]. Irrespective of the mechanisms, it is now commonly accepted that there are two types of SOTs, one is field-like (FL) and the other is damping-like (DL) [6-9]. The latter provides an efficient mechanism for switching the magnetization of FM with perpendicular magnetic anisotropy (PMA) [2,3,5,10]. In general, an assistive field parallel to the current direction (hereafter, $H_x$) is required to achieve deterministic switching[11]. As this requirement is undesirable for memory and logic applications, several approaches for achieving field-free switching have been reported[10,12-16]. Here we demonstrate that this undesirable feature of SOT-based magnetization switching can be effectively exploited for magnetic field sensor applications.

The proposed sensor, which we term it as spin torque gate (STG) sensor, can be implemented in a simple Hall bar structure made of a FM/HM bilayer with PMA. When the bilayer is driven by an ac current, the strength and polarity of $H_x$ will determine the duration within which the magnetization will stay in the up or down state, respectively, in each half-cycle of the ac current. The average anomalous Hall effect (AHE) voltage gives an output signal that is proportional to $H_x$ in the low-field range with zero offset. Since the output signal can be obtained from the average Hall voltage over many cycles, it statistically reduces the low frequency noise. The STG sensor does not require any magnetic bias which is the main cause of high cost of all types of magnetoresistance sensors including anisotropic magnetoresistance, giant magnetoresistance, and tunnel magnetoresistance sensors[17-24]. To experimentally demonstrate the STG sensor, we develop a WTe$_2$/Ti/CoFeB trilayer structure whose effective PMA can be tuned in a wide range by varying the temperature. This allows us to demonstrate proof-of-concept of operation of the STG sensor in both the deterministic and stochastic switching regimes. The sensors exhibit good linearity in the field range of ±3-10 Oe, which is almost 10 times larger than the spin Hall magnetoresistance (SMR) sensor demonstrated by us previously which uses the FL effective field as the built-in linearization mechanism[25-29]. In addition, the output voltage is also about order of magnitude larger than that of SMR sensor. Featuring the extremely simple structure, the STG sensor provides great promise with broad applications in substituting the more expensive magnetoresistance sensors.

**Operation principle of spin-torque gate magnetic field sensor**

The proposed sensor exploits the longitudinal field dependence of the magnetization reversal process of a FM/HM bilayer with PMA. Since the switching mechanism depends strongly on the size and effective PMA ($K_\perp^{eff}$) of the device, we consider the following three cases separately: 1) a single domain device with a sizable $K_\perp^{eff}$, 2) a large device which allows domain wall (DW) nucleation and propagation, and 3) a device with superparamagnetic FM layer.

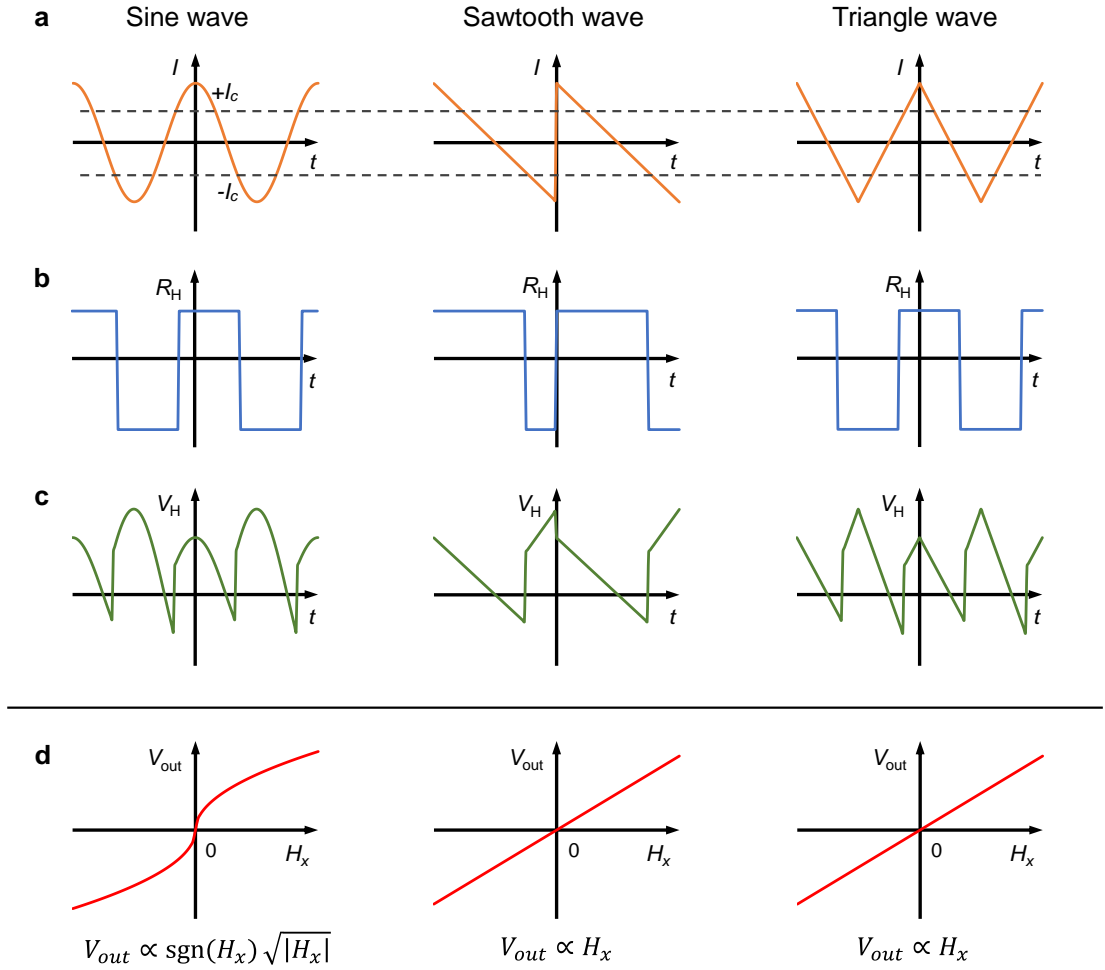

**Fig. 1 | Output voltage of STG sensor driven by different ac currents. a-c**, The current $I$, Hall resistance $R_H$, and Hall voltage $V_H$ as a function of time for sine, sawtooth and triangle waves, respectively. **d**, The output voltage $V_{out}$ as a function of in-plane field $H_x$ for different ac currents.

The SOT-induced magnetization switching in case 1) may be described by a macro-spin model[3], which predicts a critical current density of [30]

$$J_c = \frac{\sqrt{2}e}{\hbar}\frac{M_s t_{FM}}{\theta_{SH}}\left(\frac{H_k^{eff}}{\sqrt{2}} - |H_x|\right). \tag{1}$$

Here, $M_s$ ($t_{FM}$) is the saturation magnetization (thickness) of the FM layer, $H_k^{eff}$ is the effective anisotropy field, $\theta_{SH}$ is the effective spin Hall angle, $H_x$ is the longitudinal field along the current direction, $e$ is the electron charge, and $\hbar$ is the reduced Planck constant. We now consider an ac current $I(t)$ of sine, sawtooth, and triangle waveforms, passing through the FM/NM bilayer (Fig. 1a). The amplitude and period of the current are $I_0$ and $T$, respectively. The critical current $I_c$ which is the product of $J_c$ and the cross-section area of the device ($S$), is shown in Fig. 1a in dashed lines. The corresponding AHE waveform $R_H(t)$ is shown in Fig. 1b, which includes possible offset $R_0$ due to misalignment of the Hall voltage electrodes. When sketching the AHE waveform, we have assumed that it is positive (negative) when $H_x > 0$ and $I > I_c$ ($I < -I_c$), which will be reversed when $H_x$ changes sign. Figure 1c shows the Hall voltage $V_H(t) = I(t)R_H(t)$ for different ac waveforms. When $H_x \ll H_k^{eff}$, the time average of the Hall voltage, $V_{out} = \frac{1}{T}\int_0^T V_H(t)dt$, is given by

$$V_{out} = \begin{cases} \frac{\sqrt{2}R_{AHE}S}{\pi}\frac{2e}{\hbar}\frac{M_s t_{FM}}{\theta_{SH}}\text{sgn}(H_x)\sqrt{\sqrt{2}H_k^{eff}|H_x|}, & \text{sine wave} \\ \frac{R_{AHE}S}{\sqrt{2}}\frac{2e}{\hbar}\frac{M_s t_{FM}}{\theta_{SH}}H_x, & \text{sawtooth wave} \\ \frac{R_{AHE}S}{\sqrt{2}}\frac{2e}{\hbar}\frac{M_s t_{FM}}{\theta_{SH}}H_x, & \text{triangle wave} \end{cases} \tag{2}$$

where $R_{AHE}$ is the amplitude of $R_H(t)$. In all the three cases, we have set $I_0 = \frac{e}{\hbar}\frac{M_s t_{FM} S H_k^{eff}}{\theta_{SH}}$ to remove the zero-field offset. The detailed derivation of equation (2) can be found in Supplementary S1. The average output signal as a function of $H_x$ is shown in Fig. 1d, which is nonlinear for sine wave, but linear for the sawtooth and triangle waves. Equation (2) demonstrates that it is possible to realize a linear magnetic field sensor without any magnetic bias. In addition, there is also no need to use a bridge to remove the zero-field offset. As the way the signal is derived resembles the flux-gate sensor, we term it as spin torque gate (STG) magnetic field sensor. We also examined the case of square wave, but as explained in Supplementary S1, the square wave would lead to a constant amplitude of $V_{out}$ independent of $H_x$ (Supplementary Fig. 4).

The macro-spin model is valid for single domain devices. For devices with lateral dimensions larger than the DW width (case 2), the magnetization switching occurs primarily through domain wall nucleation followed by propagation, which results in a much smaller critical current density[6,11,31-41]. The DW nucleation may occur either inside the film due to defects or at the edge of the sample due to combined effect of Dzyaloshinskii-Moriya interaction (DMI), applied field and current, demagnetizing field and thermal effect, etc[6]. Latest studies suggest that DW nucleation at the edges plays a dominant role in reproducible and deterministic switching of the magnetization. According to Pizzini *et al.*[33], the nucleation at the edge may be written as $H_{n,edge} = \frac{\pi(\sigma_0 \mp 2\Delta\mu_0 M_s H_x)^2 t_{FM}}{4\mu_0 M_s p k_B T}$. Here, $\sigma_0 = 4\sqrt{AK_0} - D\pi$ is the domain wall energy density in the presence of the DMI, $D$ is the DMI constant, $K_0$ is the effective anisotropy constant, $A$ is the exchange constant, $\Delta = \sqrt{A/K_0}$, $\mu_0$ is the vacuum susceptibility, $k_B$ is the Boltzmann constant, $T$ is temperature, and $p = \ln(\frac{\tau}{\tau_0})$ with $\tau$ the waiting time and $\tau_0$ the attempt time. When $H_x \ll \frac{\sigma_0}{2\Delta\mu_0 M_s} \approx \frac{2K_0}{\mu_0 M_s}$ (when $D$ is small), one has

$$H_{n,edge} \approx \frac{4\pi A t_{FM}}{p k_B T}\left(\frac{H_k^{eff}}{2} \mp |H_x|\right). \quad (3)$$

Here, the $\mp$ sign indicates the nucleation site at opposite edges. In actual case, only the minus sign is relevant as DW propagation immediately follows its nucleation. Since $pk_BT$ can be made much larger than $4\pi A t_{FM}$, this explains why the switching current density in DW-driven switching is much smaller than the value predicted by the macro-spin model. When the nucleation field is originated from the DL effective field, the corresponding critical current density may be written as

$$J_{c,DW} = \frac{4\pi e}{\hbar}\frac{M_s A t_{FM}^2}{p\theta_{SH} k_B T}\left(\frac{H_k^{eff}}{2} - |H_x|\right). \quad (4)$$

Equation (4) is essentially the same as that of Eq. (1) except for the pre-factor. Therefore, Eq. (2) is still applicable in the incoherent switching case.

Next, we look at case 3), i.e., when the effective anisotropy becomes very small. In this case, for a single domain element, the magnetization fluctuates between $+M_z$ and $-M_z$ due to weak PMA and small thermal stability. When it is used as the free layer of magnetic tunnel junction, it results in a superparamagnetic or stochastic tunnel junction, which is an important building block for stochastic and neuromorphic computing[42-46]. Here, we explore its applications in sensors. Similar to the case of a spin with two directions pointing either parallel or anti-parallel to the external field, when a magnetic field is applied in the easy axis of superparamagnetic element with weak PMA, the average magnetization in the field direction may be written as $M_z = M_s \tanh\frac{\mu_0 M_s H_z V}{k_B T}$, and the Hall resistance is thus given by $R_H = R_0 + R_{AHE} \tanh\frac{\mu_0 M_s H_z V}{k_B T}$. Here, $H_z$ is the applied field along easy axis, $V$ is magnetic volume, and $R_0$ is the offset resistance due to voltage probe misalignment. When $H_z$ is derived from the DL SOT effective field, we have $H_z^{DL} = H^{DL} m_x$ ($H^{DL}$: magnitude of DL effective field at $m_x = 1$). When $H_x$ is small, we may write $H_z^{DL} = \left(\frac{H^{DL}}{H_k^{eff}}\right) H_x$, with $H^{DL} = \frac{\hbar}{2e}\frac{\theta_{SH}}{M_s t_{FM} S} I$. From the above relationship, we can obtain the Hall voltage at small $H_x$ as

$$V_H(t) = IR_0 + I^2 R_{AHE} \frac{\mu_0 \hbar \theta_{SH}}{2ek_B T} \frac{H_x}{H_k^{eff}}. \tag{5}$$

When the sensor is driven by a sine wave $I = I_0 \cos\omega t$, the time average of $V_H(t)$ is given by

$$V_{out} = I_0^2 R_{AHE} \frac{\mu_0 \hbar \theta_{SH}}{4ek_B T} \frac{H_x}{H_k^{eff}}, \tag{6}$$

which presents a linear relationship with $H_x$ with zero offset. In addition, a linear response can also be obtained when it is driven by other ac waveforms like square, sawtooth, and triangle waves (Supplementary S2). It is important to note that in this case, the sensor is a spin torque gated stochastic magnetic field sensor. Although stochastic magnetic field sensor has been reported before, it was implemented using a stochastic flip-flop, which requires complex circuitry to ensure linearity and accuracy[47]. In contrast, the sensor presented here is extremely simple, it comprises of only a single Hall device made of FM/HM bilayer.

**Experimental demonstration of STG sensor**

The STG sensor can be realized experimentally using any type of FM/HM bilayers or even a single layer FM as long as it has a well-defined PMA, sizable damping-like SOT, and the requirement of a longitudinal field for deterministic switching. We chose to use WTe$_2$ (5)/Ti (2)/CoFeB (1.5)/MgO (2)/Ta (1.5) (the number in the parentheses indicates the layer thickness in nm) for experimental implementation of the STG sensor because WTe$_2$ is reported to have a large charge-spin conversion efficiency[48-55]. Moreover, recently two groups have observed evident SOT in WTe$_2$ thin films prepared by sputtering[53,54], which is of more practical relevance when it comes to device applications. The CoFeB layer exhibits well-defined PMA in the multilayer structure, which arises mainly from the interface anisotropy of CoFeB/MgO [56], and the 2 nm Ti insertion layer functions as a seed layer to promote the PMA [57,58]. The other consideration for using this structure is that its coercivity varies strongly with temperature which, in addition to the STG sensor, also allows to realize spin torque gated stochastic magnetic sensor when the effective PMA becomes very weak.

Smooth WTe$_2$ thin films with the root mean square (rms) roughness of ~0.2 nm were obtained by process optimization (Supplementary S3). The resistivity of the WTe$_2$ thin film at room temperature is around $8.7\times10^3$ μΩ cm, which is two orders of magnitude larger than that of normal metals. The increase of resistivity upon decreasing the temperature indicates a non-metallic behaviour of sputtered WTe$_2$ (Supplementary S4). The highly resistive WTe$_2$ thin film is advantageous for sensor application since a larger anomalous Hall resistance will be obtained due to less current shunting in WTe$_2$. We further characterized the SOT in WTe$_2$-based ferromagnetic heterostructures by using the harmonic Hall measurements (Supplementary S5). The WTe$_2$ thickness dependence study of SOT in WTe$_2$ ($t_{WTe_2}$)/CoFeB was conducted, from which the DL SOT efficiency is found to peak at around $t_{WTe_2} =$ 5 nm with a value of $0.67\times10^5$ Ω$^{-1}$ m$^{-1}$ (Supplementary S6). Only a slight decrease of the SOT was observed after a thin Ti layer is inserted between WTe$_2$ and CoFeB, which suggests the high spin transparency of Ti [57-59] (Supplementary S7 and S8). With the current distribution in each layer considered, the DL SOT field for the WTe$_2$ (5)/Ti (2)/CoFeB (1.5) stack is estimated to be 26.3 Oe per $10^7$ A/cm$^2$.

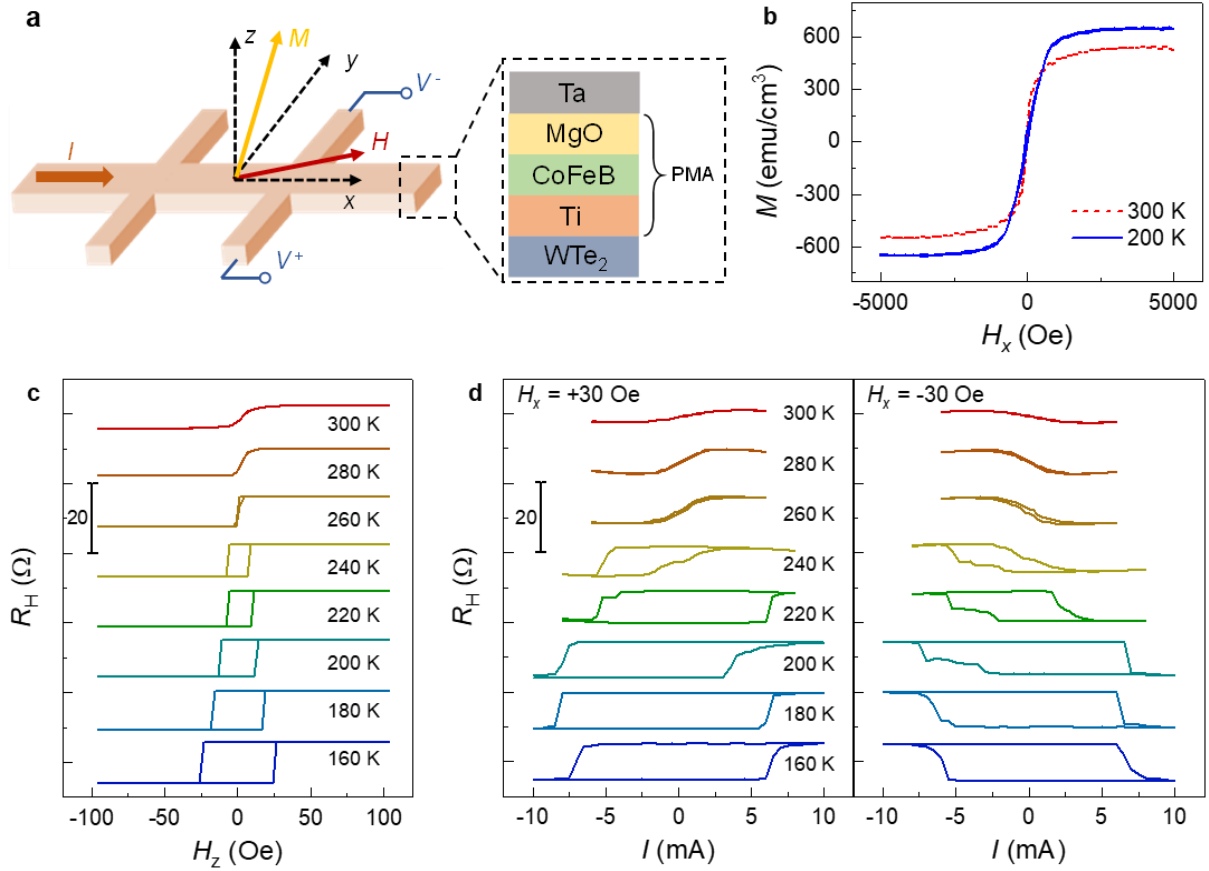

**Fig. 2 | Current-induced switching of WTe$_2$/Ti/CoFeB. a**, Schematic of the Hall bar device and electrical measurement configuration. **b**, $M-H$ loops of WTe$_2$ (5)/Ti (2)/CoFeB (1.5) coupon film at 200 K and 300 K, respectively, with the field swept in *x*-direction. **c**, AHE curves of WTe$_2$ (5)/Ti (2)/CoFeB (1.5) measured with a perpendicular field $H_z$ and a small dc current of 200 µA at different temperatures from 160 K to 300 K. **d**, Current-induced switching loops of WTe$_2$ (5)/Ti (2)/CoFeB (1.5) at different temperatures with an in-plane assistive field $H_x$ of +30 Oe and -30 Oe, respectively. The pulse width of the pulse current is fixed at 2 ms.

Figure 2a shows the schematic of a Hall bar device used for sensor demonstration. In order to estimate the effective magnetic anisotropy, we measured the $M-H$ curve of WTe$_2$ (5)/Ti (2)/CoFeB (1.5) coupon film with an in-plane field (Fig. 2b), from which the $M_s$ and $H_k^{eff}$ are obtained as 550 emu/cm$^3$ and 800 Oe at room temperature and 650 emu/cm$^3$ and 1200 Oe at 200 K, respectively. This gives a thermal stability factor of around 10 and 27 at room temperature and 200 K, respectively (assuming the domain size is 40 nm). This allows us to demonstrate proof-of-concept STG sensor in both the DW-driven (low temperature) and stochastic (high temperature) region using the same device. Figure 2c shows the AHE curves at different temperatures from 160 K to 300 K. As can be seen, the device exhibits well-defined PMA at low temperature and superparamagnetic behavior near room temperature with the coercivity ranging from 0 Oe to 25 Oe. Furthermore, Fig. 2d shows the current-induced switching loops of the same sample at different temperatures with $H_x$ of ±30 Oe. As can be seen, the AHE curve corroborates well the DL SOT-induced switching mechanism, with the switching polarity determined by both the current and $H_x$ directions. Additionally, the opposite switching polarity in WTe$_2$/Ti/CoFeB as compared to Pt/Co[11,41,60] indicates a negative sign of spin Hall angle in WTe$_2$, which is consistent with the results of SOT measurements (Supplementary S5 and S6). The switching current $I_{sw}$ at $H_x = \pm 30$ Oe is around 8 mA (3 mA) at 160 K (300 K), corresponding to a current density of 1.33×10$^7$ A/cm$^2$ (5×10$^6$ A/cm$^2$). At low temperature, the switching is quite abrupt at the initial phase, but it becomes gradual at the final phase, indicating that the switching is dominated by DW nucleation and propagation/expansion from the nucleation sites. The DW-dominated switching mechanism becomes more apparent at intermediate temperatures, as manifested in the multiple-step switching. The irregular hysteresis loop suggests that the switching evolves from DW-driven to stochastic mode around 240 K. This unique characteristic of WTe$_2$ (5)/Ti (2)/CoFeB (1.5) facilitates the demonstration of STG sensor at low temperature and spin torque gated stochastic sensor at room temperature.

To demonstrate the proof-of-concept operation of the STG sensor, we apply an ac current with triangle waveform to the device at 160 K and measure the time average Hall voltage as a function of $H_x$. The amplitude and frequency of the applied ac current are 11.5 mA and 115 Hz, respectively. The Hall voltage was measured by a Data Acquisition (DAQ) device (Supplementary S10), with the output voltage $V_{out}$ at each field averaged over 2×10$^6$ sampling points for a duration of 2 s. Figure 3a shows the time averaged $V_{out}$ in response to $H_x$ swept forth and back between -3 Oe and 3 Oe. A linear response with < 2.5% linearity error (as shown in lower right inset), nearly zero dc offset, and negligible hysteresis is obtained. The upper left inset in Fig. 3a shows $V_{out} - H_x$ dependence in a larger range from -10 Oe to 10 Oe. It is worth pointing out that the data shown in Fig. 3a are raw data without any dc offset compensation. Both the zero dc offset and small hysteresis are the unique features of the STG sensor despite the fact that the sensor element has a finite coercivity and there is always a dc offset in the AHE signal due to misalignment of the Hall voltage probes. Similar results were also obtained using the sawtooth wave as the driving current (Supplementary S11).

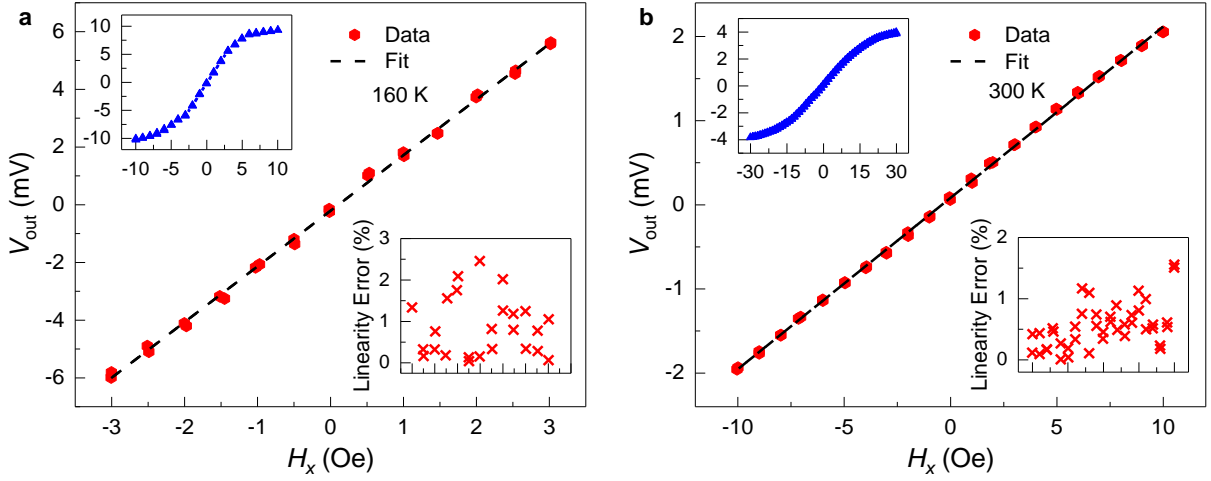

**Fig. 3 | Experimental demonstration of STG magnetic sensor. a,** Output voltage $V_{out}$ in response to an external field $H_x$ swept forth and back between -3 Oe and 3 Oe at 160 K (dashed line is the guide for the eye). **b,** $V_{out}$ in response to $H_x$ swept forth and back between -10 Oe and 10 Oe at 300 K (dashed line is the guide for the eye). Upper left inset in **a, b**: the output response curve in a larger sweeping range. Lower right inset in **a, b**: linearity error within the sweeping field range.

Next we turn to the performance of the sensor at room temperature. As shown in Fig. 2c, the hysteresis of this sample almost disappears when $T > 260$ K, though it still shows PMA. Since the lateral size of the active element of the sensor is 15 $\mu$m × 5 $\mu$m, we may reasonably postulate that there are finite number of superparamagnetic elements inside the region which contribute to the AHE signal. If the dispersion of effective anisotropy is not that large, the overall $M_z$ dependence on temperature should still follow the $M_z = M_s \tanh \frac{\mu_0 M_s H_z V}{k_B T}$ relation. The $R_H - I$ curves at 300 K can be well fitted using the hyperbolic tangent function (Supplementary S12). Therefore, we shall be able to verify the proposed spin torque gated stochastic sensor using the same WTe$_2$ (5)/Ti (2)/CoFeB (1.5) device at room temperature.

Based on aforementioned discussion, in this case the amplitude of the ac current is not required to be larger than the critical current due to the absence of hysteresis. Figure 3b shows $V_{out}$ of the sensor in response to $H_x$ swept forth and back between -10 Oe and 10 Oe driven by a square wave with an amplitude of 2 mA and frequency of 115 Hz. The upper left inset displays the output response curve in a larger sweeping range (-30 Oe to 30 Oe). The linearity error within the field range of ±10 Oe is less than 2% as shown in the lower right inset, indicating a good linearity of the sensor. The sensitivity of this device at room temperature is 102 mΩ/Oe, which can be further improved by optimizing the materials, layer structure, and dimension of the sensor. One should note that, although we detect the readout signal using AHE in this work, the basic operation principle can also be applied to three-terminal sensors in which the readout can be from a magnetic tunnel junction. In this case, the signal amplitude can be increased by more than two orders.

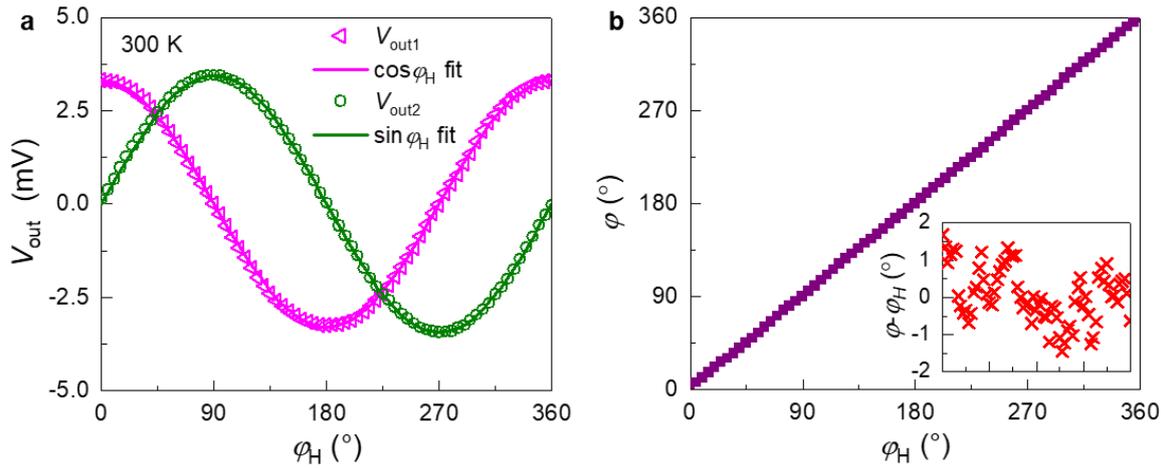

**Fig. 4 | Proof-of-concept application of STG sensor in angle detection. a**, Output voltage of the Hall device with $\varphi_H = 0°$ aligned along $x$ and $y$ direction, respectively, and a field of 20 Oe is rotated in the film plane. **b**, Measured rotational angle $\varphi$ as a function of the $\varphi_H$. Inset: angle error $\varphi - \varphi_H$ in the full range of 360°.

Due to its simple structure, good linearity, and negligible offset, the STG sensor is expected to have many potential applications. As one of the proof-of-concept experiments, here we demonstrate its application in position sensing. To this end, we apply a rotational field of 20 Oe in the $xy$ plane and measure the output of the sensor at each position. Figure 4a shows the output voltage $V_{out1}$ and $V_{out2}$ of the sensor with the longitudinal direction at $\varphi_H = 0°$ placed along $x$ direction and $y$ direction, respectively, as a function of the external field angle $\varphi_H$. The sensor was driven by a square wave with the amplitude of 2 mA and frequency of 115 Hz. As can be seen, $V_{out1}$ and $V_{out2}$ have cosine and sine dependence on the field angle, which can be well fitted by $V_1 \cos\varphi_H$ and $V_2 \sin\varphi_H$, respectively, where $V_1$ and $V_2$ are the amplitudes of $V_{out1}$ and $V_{out2}$. The field angle $\varphi$ can be calculated using the equation:

$$\varphi = \text{acrtan2}\left[-\frac{V_{out2}}{V_2}, -\frac{V_{out1}}{V_1}\right] + \pi. \tag{7}$$

The detected angle $\varphi$ from the above equation is shown in Fig. 4b versus the actual angle $\varphi_H$, from which a very linear relationship of $\varphi$ and $\varphi_H$ is observed. The calculated angle error $(\varphi - \varphi_H)$ at each position is shown in the inset of Fig. 4b. As can be seen, the angle error is mostly within 1° except for a few positions, with an average and maximum value of 0.59° and 1.68°, respectively, indicating the accurate angle detection ability of the STG sensor. Comparing with the SOT-enabled angular position sensor that senses second harmonic Hall voltage reported in our previous work[61], the output voltage of the STG sensor is more than one order of magnitude larger, and in addition, it does not suffer from the influence of dc offset and planar Hall effect.

In conclusion, we have proposed and demonstrated experimentally a spin torque gate magnetic sensor based on the longitudinal field dependence of current-induced magnetization switching in FM/NM bilayers. Driven by an ac current, and with the Hall voltage averaged over a large number of cycles, the sensor exhibits a linear response to the external magnetic field with negligible hysteresis and dc offset. As a proof-of-concept demonstration, we implemented the STG sensor on the Hall bar device with $WTe_2$/Ti/CoFeB stacks, where sputtering deposited $WTe_2$ was employed as the SOT generator due to its large spin Hall angle. There is no doubt that the performance of the sensor can be further improved through materials and design optimization. The extremely simple structure of the STG sensor makes it attractive for many potential applications, as manifested in our proof-of-principle experiments in angle detection.

**Methods**

All the materials except for Ti were deposited on SiO$_2$ (300 nm)/Si substrates using magnetron sputtering with a base and working pressure of <1×10$^{-8}$ and 1.5 - 3×10$^{-3}$ Torr, respectively. The Ti layer was deposited by e-beam evaporation in the same vacuum system with the sputter without breaking the vacuum. Standard photolithography and liftoff techniques were used to fabricate the Hall bars. The Hall bar length, width, spacing between voltage electrodes and width of voltage electrode are 120 μm, 15 μm, 30 μm, and 5 μm, respectively. The Mircotech laserwriter system with a 405 nm laser was used to directly expose the photoresist (Mircoposit S1805), after which it was developed in MF319 to form the Hall bar pattern. After film deposition, the photoresist was removed by a mixture of PG remover and acetone to complete the Hall bar fabrication.

The surface roughness and sputtering rate of thin films were characterized using a Veeco Dimension 3100 AFM system. The magnetic properties were characterized using Quantum Design MPMS3, with the resolution of <1×10$^{-8}$ emu. The electrical measurements were performed in the Quantum Design VersaLab PPMS with a sample rotator. The ac/dc current is applied by the Keithley 6221 current source. The longitudinal or Hall voltage is measured by the Keithley 2182 nanovoltmeter (for dc voltage), or the 500 kHz MFLI lock-in amplifier from Zurich Instruments (for harmonic voltage), or the National Instruments Data Acquisition (NI-DAQ) device (for acquiring the voltage with a large sampling rate of 1 MHz).

**Acknowledgements**

This project is supported by the Ministry of Education, Singapore under its Tier 2 Grants (Grants No. MOE2018-T2-1-076 and No. MOE2017-T2-2-011).

**Author contributions**

Y.H.W. conceived the idea and supervised the project. H.X. and Y.H.W. designed the experiments. H.X. and X.C. fabricated the samples. H.X. and Z.Y.L. set up the electrical measurements. H.X. performed the AFM characterizations, the electrical and magnetic measurements. Y.H.W., H.X. and Z.Y.L. analyzed the data. All authors discussed the results. H.X. and Y.H.W. wrote the manuscript and all the authors contributed to the final version of manuscript.

**Competing Interests**

The authors declare no competing interests.